\documentclass[10pt]{article}
\pdfoutput=1
\usepackage[utf8x]{inputenc}
\usepackage{amssymb,amsmath,mathrsfs,enumerate}
\usepackage{graphicx,rotate,multicol}
\usepackage{wrapfig}
\usepackage[colorlinks=true,
linkcolor=red,
urlcolor=blue,
citecolor=blue]{hyperref}
\usepackage{color}
\usepackage{soul}
\usepackage{cite}
\usepackage{bbold}
\usepackage{verbatim}
\usepackage[symbol]{footmisc}
\long\def\rpl#1!!#2!!{\textcolor{red}{#1} \textcolor{blue}{#2}}

\def \order(#1){{\cal O} \left(#1 \right)}

\evensidemargin=\oddsidemargin
\def\Eqn#1{Eq.\ (\ref{#1})}
\def\Eqs#1#2{Eqs.\ (\ref{#1}) and (\ref{#2})}

\begin{document}

\begin{flushright}
	LU TP 19-23\\
	IPMU19-0052
\end{flushright}

\begin{center}
	{\Large \bf Alignment limit in three Higgs-doublet models } \\
	\vspace*{1cm} {\sf Dipankar
          Das$^{a,}$\footnote[1]{dipankar.das@thep.lu.se},~Ipsita
          Saha$^{b,}$\footnote[2]{ipsita.saha@ipmu.jp}} \\
	\vspace{10pt} {\small \em 
          $^a$Department of Astronomy and Theoretical Physics, Lund University, S\"{o}lvegatan 14A, Lund 22362, Sweden \\
		$^b$Kavli IPMU (WPI), UTIAS, University of Tokyo, Kashiwa, 277-8583, Japan}
	
	\normalsize
\end{center}

\renewcommand*{\thefootnote}{\arabic{footnote}}
\setcounter{footnote}{0} 
\begin{abstract}
The LHC Higgs data is showing a gradual inclination towards the SM result and realization of an SM-like limit
becomes essential for BSM scenarios to survive. Considering the accuracy that can be achieved in future colliders,
BSMs that acquire the alignment limit with an SM-like Higgs boson can surpass others in the long run.
Using a convenient parametrization, we demonstrate that the
alignment limit for CP-conserving 3HDMs takes on the same analytic
structure as that in the case of 2HDMs. Using the example of
a $Z_3$-symmetric 3HDM, we illustrate how such alignment conditions
can be efficiently implemented for numerical analysis in a
realistic scenario.
\end{abstract}

\bigskip

\section{Introduction} \label{s:intro}
In the post Higgs discovery era, the absence of any direct sign of 
new physics~(NP) at the LHC has already pushed many beyond the Standard Model~(BSM) scenarios at bay. An alternative way to find hints for NP
will be to look for deviations of Higgs couplings from their
corresponding Standard Model~(SM) predictions. However, since the
discovery of the Higgs boson, the LHC Higgs data has been gradually
drifting towards the SM expectations. In fact, deviations in the
observed Higgs signal strengths from their respective SM values seem
to have been significantly reduced with the increased sensitivity at the LHC Run II~\cite{CMS:2018lkl,ATLAS:2018doi}. But it is still
possible for BSM scenarios to be hiding behind the curtain,
camouflaging themselves with an SM-like Higgs. Thus, in anticipation
that the LHC Higgs data will continue to incline towards the SM
expectations with increasing accuracy, those BSM scenarios which
can deliver an SM-like Higgs in a certain {\em alignment limit}
will have an upper hand in the future survival race. In this paper,
we uphold the three Higgs-doublet models~(3HDMs) as potential
candidates for such BSM scenarios.
Adding replicas of the SM Higgs-doublet constitutes one of the
simplest ways to extend the SM for such extensions do not alter
the tree-level value of the electroweak~(EW) $\rho$-parameter.
A lot of attention has already been given to the two Higgs-doublet
models~(2HDMs)\cite{Branco:2011iw,Bhattacharyya:2015nca} where the scalar sector of the SM is extended
by an additional Higgs-doublet. As a next step, recent years have
seen a growing interest in the topic of 3HDMs\cite{Ferreira:2008zy,Machado:2010uc,Aranda:2012bv,Ivanov:2012ry,Ivanov:2012fp,Felipe:2013ie,Felipe:2013vwa,Keus:2013hya,Aranda:2013kq,Ivanov:2014doa,Das:2014fea,Maniatis:2014oza,Das:2015gyv,Maniatis:2015kma,Moretti:2015cwa,Chakrabarty:2015kmt,Merchand:2016ldu,Emmanuel-Costa:2016vej,Yagyu:2016whx,Bento:2017eti,Emmanuel-Costa:2017bti,Camargo-Molina:2017klw,Pramanick:2017wry,deMedeirosVarzielas:2019rrp} where 
two additional Higgs-doublets are added to the SM scalar sector.
Therefore, 3HDMs conform to the aesthetic appeal of having three
generations of scalars on equal footing to three 
fermionic generations in
the SM\cite{Das:2015sca}. In such models, the $125$~GeV scalar observed at the
LHC is only the first to appear in a series of many others to
follow. Evidently, the rich scalar spectrum of the 3HDM must
contain one physical scalar having properties similar to those
of the SM Higgs boson, which can serve as a competent candidate
for the $125$~GeV scalar. The limit in which the lightest CP-even
scalar possesses SM-like tree-level couplings with the fermions
and the vector bosons is usually dubbed as the alignment limit.
In the case of 2HDMs, the analytic condition for alignment is
well known\cite{Gunion:2002zf,Carena:2013ooa,Dev:2014yca} and it has been very useful in analyzing
2HDMs in the light of the Higgs data\cite{Craig:2012vn,Bhattacharyya:2013rya,Bhattacharyya:2014nja,Das:2015qva,Das:2015mwa,Das:2015zwa,Bernon:2015qea,Bernon:2015wef,Das:2017zrm,Grzadkowski:2018ohf}. However, in the
case of multi Higgs-doublet models with more than two Higgs-doublets,
although the general recipe for obtaining alignment has been
studied earlier in the literature\cite{Das:2015gyv,Pilaftsis:2016erj,Bento:2017eti}, analytic expressions
suitable for practical use are currently lacking. In this paper,
we attempt to find the conditions for alignment in 3HDMs, in a
simple form that can be easily implemented in practical models
to investigate several aspects of 3HDMs. In fact, using a convenient
parametrization for CP conserving 3HDMs, we will demonstrate that
the requirement of an SM-like Higgs results in simple equations
which resemble very much to the alignment condition in CP
conserving 2HDMs.

Our paper will be organized as follows. In Sec.~\ref{s:a-limit}
we will briefly revisit the general prescription for obtaining
an SM-like Higgs in multi Higgs-doublet models. Then we will use
this to recover the alignment limit in 2HDMs, and extend the idea
to the case of 3HDMs. In Sec.~\ref{s:example} we will illustrate
how our results of Sec.~\ref{s:a-limit} can substantially
simplify the analysis of a CP conserving 3HDM. Finally, our
findings will be summarized in Sec.~\ref{s:summary}.

%
\section{The alignment limit } 
\label{s:a-limit} 
As mentioned earlier, the alignment limit is defined as the set
of conditions under which the lightest CP-even scalar mimics the
SM Higgs by possessing SM-like gauge and Yukawa couplings at the
tree-level. To illustrate how such a limit can be reached in a
multi Higgs-doublet scenario, let us consider a general $n$
Higgs-doublet model~($n$HDM) where the $k$-th doublet is
expanded in terms of its component fields as follows:
\begin{eqnarray}
\label{e:field_definition}
\phi_k = \begin{pmatrix}
w_k^+ \\
(h_k+ i z_k)/\sqrt{2}
\end{pmatrix} \,, \qquad (k=1,2,\dots ,n) \,.
\end{eqnarray}
Under the assumption that all the parameters in the $n$HDM scalar
potential are real, there will be no mass mixing between the $h_k$
and the $z_k$ fields. Denoting by $\left\langle \phi_k\right\rangle  = 
v_k/\sqrt{2}$ the vacuum expectation value~(VEV) for $\phi_k$ after
spontaneous symmetry breaking~(SSB), the total EW VEV, $v$, can be
identified as
\begin{eqnarray}
v^2 = \sum_{k=1}^{n} v_k^2 =  (246 \,\, {\rm GeV})^2 \,.
\end{eqnarray}
To gain some intuitive insights into the alignment limit of an
$n$HDM, it is instructive to take a closer look at the scalar
kinetic Lagrangian which contains the following trilinear
couplings:
\begin{eqnarray}
{\mathscr L}_{\rm kin}^S = \sum_{k=1}^n |D_\mu \phi_k|^2 \ni
\frac{g^2}{2} W_\mu^+ W^{\mu -} \left(\sum_{k=1}^n v_kh_k \right)
\equiv \frac{g^2v}{2} W_\mu^+ W^{\mu -} \left(\frac{1}{v}
\sum_{k=1}^n v_kh_k \right) \,,
\end{eqnarray}
where $g$ stands for the $SU(2)_L$ gauge coupling. Clearly, the
combination,
\begin{eqnarray}
\label{e:H0}
H_0 = \frac{1}{v} \sum_{k=1}^n v_k h_k \,,
\end{eqnarray}
will resemble to the SM Higgs boson in its tree-level gauge
couplings. It is also not very difficult to show that $H_0$ will
have SM-like Yukawa couplings too\cite{Das:2015gyv}.
However, this state $H_0$, in general, is not guaranteed to be
a physical eigenstate. Therefore, the alignment limit will
emerge as the limit when $H_0$ aligns itself completely with
the lightest CP-even physical scalar~($h$) in the spectrum.
%

\subsection{Alignment in 2HDM}
To begin with let us apply the result of the previous section to
retrieve the alignment limit in the 2HDM case. Following the
definition in \Eqn{e:H0}, the state $H_0$ and its orthogonal
combination, $R$, can be obtained by the following orthogonal
rotation:
\begin{eqnarray}
\label{e:Ob2hdm}
    \begin{pmatrix} H_0 \\ R \end{pmatrix} = {\cal O}_\beta
    \begin{pmatrix} h_1 \\ h_2 \end{pmatrix} =
    \begin{pmatrix} \cos\beta & \sin\beta \\ 
    -\sin\beta & \cos\beta \end{pmatrix}
    \begin{pmatrix} h_1 \\ h_2 \end{pmatrix}  \,,
\end{eqnarray}
where $\tan\beta=v_2/v_1$. On the other hand, the physical mass
eigenstates, $h$ and $H$, are extracted using another orthogonal rotation
characterized by the angle, $\alpha$, as follows:
\begin{eqnarray}
\label{e:Oa2hdm}
\begin{pmatrix} h \\ H \end{pmatrix} = {\cal O}_\alpha
\begin{pmatrix} h_1 \\ h_2 \end{pmatrix} =
\begin{pmatrix} \cos\alpha & \sin\alpha \\ 
-\sin\alpha & \cos\alpha \end{pmatrix}
\begin{pmatrix} h_1 \\ h_2 \end{pmatrix}  \,.
\end{eqnarray}
Inverting \Eqn{e:Ob2hdm} and then plugging it on the right hand
side of \Eqn{e:Oa2hdm}, we can write
\begin{eqnarray}
\label{e:ab}
\begin{pmatrix} h \\ H \end{pmatrix} = {\cal O}_\alpha {\cal O}_\beta^T
\begin{pmatrix} H_0 \\ R \end{pmatrix} =
\begin{pmatrix} \cos(\alpha-\beta) & \sin(\alpha-\beta) \\ 
-\sin(\alpha-\beta) & \cos(\alpha-\beta) \end{pmatrix}
\begin{pmatrix} H_0 \\ R \end{pmatrix}  \,.
\end{eqnarray}
Thus, $h$ will completely overlap with $H_0$ if
\begin{eqnarray}
    \cos(\alpha-\beta) = 1  &\Rightarrow& \alpha=\beta \,,
\end{eqnarray}
which defines the alignment limit in 2HDMs.\footnote{In a more
    conventional set up, the definition of $\alpha$ differs
    from our definition by $\pi/2$ so that the alignment
    condition reads $\cos(\alpha-\beta)=0$.}

\subsection{Alignment in 3HDM}
In the case of 3HDMs let us first parametrize the VEVs as follows:
\begin{eqnarray}
\label{e:VEVs}
    v_1 = v \cos \beta_1 \cos\beta_2\,, \qquad
    v_2 = v \sin\beta_1 \cos\beta_2\,,  \qquad
    v_3 = v \sin\beta_2 \,.
\end{eqnarray}
Thus, the analogue of \Eqn{e:Ob2hdm} for 3HDM will read
\begin{eqnarray}
\label{e:Ob}
\begin{pmatrix} H_0 \\ R_1 \\ R_2 \end{pmatrix} = {\cal O}_\beta
\begin{pmatrix} h_1 \\ h_2 \\ h_3 \end{pmatrix} =
\begin{pmatrix} \cos\beta_2 \cos\beta_1 & \cos\beta_2 \sin\beta_1 
&  \sin\beta_2 \\ 
-\sin\beta_1 & \cos\beta_1  &  0  \\
-\cos\beta_1 \sin\beta_2 & -\sin\beta_1\sin\beta_2 & \cos\beta_2
\end{pmatrix}
\begin{pmatrix} h_1 \\ h_2 \\ h_3 \end{pmatrix}  \,.
\end{eqnarray}
Note that the first row of ${\cal O}_\beta$ in the above equation
is motivated from \Eqn{e:H0} but the choices for the second and the
third rows are not unique. Our analysis does not depend on these
choices. Next, in analogy with \Eqn{e:Oa2hdm}, 
${\cal O}_\alpha$ will now be a $3\times 3$ orthogonal matrix
which takes us to the physical basis,
$(h \, H_1 \, H_2)^T$. Therefore, we can decompose
${\cal O}_\alpha$ as follows:
\begin{subequations}
\label{e:Oa}
\begin{eqnarray}
{\cal O}_\alpha &=& {\cal R}_3 \cdot  {\cal R}_2\cdot {\cal R}_1 \,,
\end{eqnarray}
where,
\begin{eqnarray}
\label{e:R}
{\cal R}_1 = \left(\begin{array}{ccc}
\cos \alpha_1 & \sin \alpha_1 & 0 \\
-\sin \alpha_1 & \cos \alpha_1 & 0 \\
0 & 0 & 1 \\
\end{array}\right)\,, \quad {\cal R}_2 = \left(\begin{array}{ccc}
\cos \alpha_2 & 0 & \sin \alpha_2  \\
0 & 1 & 0 \\
-\sin \alpha_2 & 0 & \cos \alpha_2 \\
\end{array}\right)\,,  \quad
{\cal R}_3 = \left(\begin{array}{ccc}
1 & 0 & 0 \\
0 & \cos \alpha_3 &  \sin \alpha_3  \\
0 & -\sin \alpha_3 & \cos \alpha_3 \\
\end{array}\right)\,.
\end{eqnarray}
\end{subequations}
Now, similar to \Eqn{e:ab}, we can write
\begin{eqnarray}
\label{e:ab3hdm}
\left(\begin{array}{c}
h \\
H_1 \\
H_2 \\
\end{array}\right)
&=&{\cal O}_ \alpha \cdot {\cal O}_\beta^T
\left(\begin{array}{c}
H^0 \\
R_1 \\
R_2 \\
\end{array}\right)
\end{eqnarray}
in case of 3HDMs. For the convenience of notations
in our analysis of 3HDMs, we introduce the matrix
\begin{eqnarray}
\label{e:O}
{\cal O}\equiv{\cal O}_ \alpha \cdot {\cal O}_\beta^T \,,
\end{eqnarray}
where ${\cal O}_\beta$ and ${\cal O}_\alpha$ have been defined
in \Eqs{e:Ob}{e:Oa} respectively. Thus, for $h$ to overlap
completely with $H_0$, we must require\footnote{Note that,
    \Eqn{e:O11} will automatically ensure ${\cal O}_{12}=
    {\cal O}_{21}={\cal O}_{13}={\cal O}_{31}=0$ due to the
    orthogonality of ${\cal O}$.}
\begin{eqnarray}
\label{e:O11}
    {\cal O}_{11} = 1 \,,
\end{eqnarray}
which can be expressed as
\begin{eqnarray}
\label{e:al1}
    \cos\alpha_2 \cos\beta_2 \cos(\alpha_1-\beta_1)
    +\sin\alpha_2 \sin\beta_2 = 1 \,.
\end{eqnarray}
After some simple trigonometric manipulations the above condition
can be recast in the following form:
\begin{eqnarray}
\label{e:al2}
    \left[\sin\left(\frac{\alpha_1-\beta_1}{2}\right)
    \cos\left(\frac{\alpha_2+\beta_2}{2}\right)\right]^2
    +\left[\cos\left(\frac{\alpha_1-\beta_1}{2}\right)
    \sin\left(\frac{\alpha_2-\beta_2}{2}\right)\right]^2
    = 0 \,,
\end{eqnarray}
which implies
\begin{subequations}
\label{e:al3}
\begin{eqnarray}
 && \sin\left(\frac{\alpha_1-\beta_1}{2}\right)
\cos\left(\frac{\alpha_2+\beta_2}{2}\right) = 0 \,, \\
{\rm and,} &&
\cos\left(\frac{\alpha_1-\beta_1}{2}\right)
\sin\left(\frac{\alpha_2-\beta_2}{2}\right) =0 \,.
\end{eqnarray}
\end{subequations}
These conditions together define the alignment limit for a
CP-conserving 3HDM. One can easily check that the conditions of
\Eqn{e:al3} admit the following two possibilities:
\begin{eqnarray}
    \alpha_1=\beta_1 \,; &&  \alpha_2=\beta_2 \,,
    \label{e:alfinal} \\
    {\rm or,} \qquad \alpha_1=\pi +\beta_1 \,; &&  
    \alpha_2=\pi-\beta_2 \,.
    \label{e:alalt}
\end{eqnarray}
But, using \Eqn{e:O} it can be verified that choosing condition
(\ref{e:alalt}) instead of condition~(\ref{e:alfinal}) only amounts
to redefinitions of the physical fields $H_1$ and $H_2$ as
\begin{eqnarray}
    H_1 \to -H_1 \,, &{\rm and}& H_2 \to -H_2 \,,
\end{eqnarray}
which are physically equivalent. Therefore, we choose
condition~(\ref{e:alfinal}) as the definition of alignment
limit in 3HDMs, which, when compared with \Eqn{e:ab}, looks
very similar to the 2HDM case.

In passing, we note that \Eqn{e:al1} can be trivially satisfied
in the limit $\sin\alpha_2 \approx \sin\beta_2 \approx 1$
which, in view of \Eqn{e:VEVs}, corresponds to the situation
where $\phi_3$ acquires the entire EW VEV and consequently
$\phi_1$ and $\phi_2$ are rendered (almost) inert. However,
barring such extreme VEV hierarchies, \Eqn{e:alfinal} must
be obeyed so that an SM-like Higgs may emerge from the 3HDM
scalar spectrum.


\section{An example: 3HDM with $Z_3$ symmetry}
\label{s:example}
At this point, it is reasonable to ask how close we need to be
to the alignment limit in view of the current Higgs data. To
analyze this, we proceed by defining the Higgs coupling modifiers
as
\begin{eqnarray}
    \kappa_x = \frac{g_{hxx}}{(g_{hxx})^{\rm SM}} \,,
\end{eqnarray}
where $x$ stands for the massive fermions and vector bosons.
Keeping in mind that among $H_0$, $R_1$ and $R_2$ in
\Eqn{e:ab3hdm}, only $H_0$ possesses trilinear coupling
of the form $H_0VV$ ($V=W,Z$), we conclude
\begin{eqnarray}
    \kappa_V \equiv {\cal O}_{11} =
    \cos\alpha_2 \cos\beta_2 \cos(\alpha_1-\beta_1)
    +\sin\alpha_2 \sin\beta_2 \,.
\end{eqnarray}
To obtain the fermionic coupling modifiers we need to know how
the Higgs-doublets couple to the fermions. For this, we consider
the example of a $Z_3$ symmetric 3HDM, in which the scalar
doublets $\phi_1$ and $\phi_2$ transform nontrivially
as follows:
\begin{eqnarray}
    \phi_1 \to \omega\, \phi_1 \,, \qquad
    \phi_2 \to \omega^2 \phi_2 \,,
\end{eqnarray}
where $\omega=e^{2\pi i/3}$. Furthermore, some of the right
handed fermionic fields transform under $Z_3$ as follows:
\begin{eqnarray}
d_R \to \omega\, d_R \,, \qquad
\ell_R \to \omega^2 \ell_R \,,
\end{eqnarray}
where $d_R$ and $\ell_R$ denote the right handed down type quarks
and charged leptons respectively. The rest of the fields in the
theory are assumed to remain unaffected under $Z_3$. With these
charge assignments, $\phi_3$ and $\phi_2$ will be responsible
for masses of the up and down type quarks respectively,
whereas $\phi_1$ will give masses to the charged leptons.
Consequently, the fermionic coupling modifiers will be given
by
\begin{subequations}
    \begin{eqnarray}
  \kappa_u &=& \frac{\sin\alpha_2}{\sin\beta_2} \,, \\
  \kappa_d &=& \frac{\sin\alpha_1\cos\alpha_2}{\sin\beta_1\cos\beta_2}
  \,, \\
  \kappa_\ell &=& \frac{\cos\alpha_1\cos\alpha_2}{\cos\beta_1\cos\beta_2}
  \,,
    \end{eqnarray}
\end{subequations}
all of which, as expected, approaches unity in the alignment
limit defined by \Eqn{e:alfinal}.
\begin{figure}[ht]
	\begin{minipage}{0.46\textwidth}
		\centerline{\includegraphics[scale=0.3]{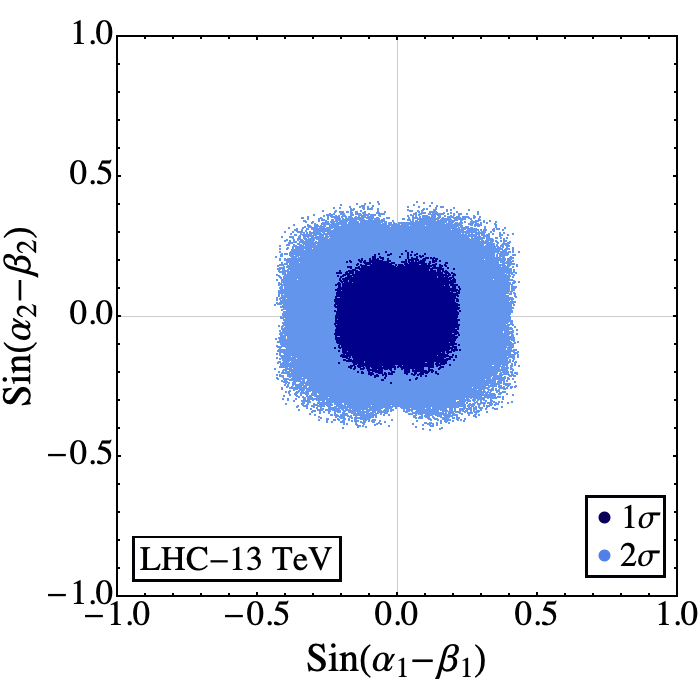}}
	\end{minipage}
	\hfill
	\begin{minipage}{0.46\textwidth}
		\centerline{\includegraphics[scale=0.3]{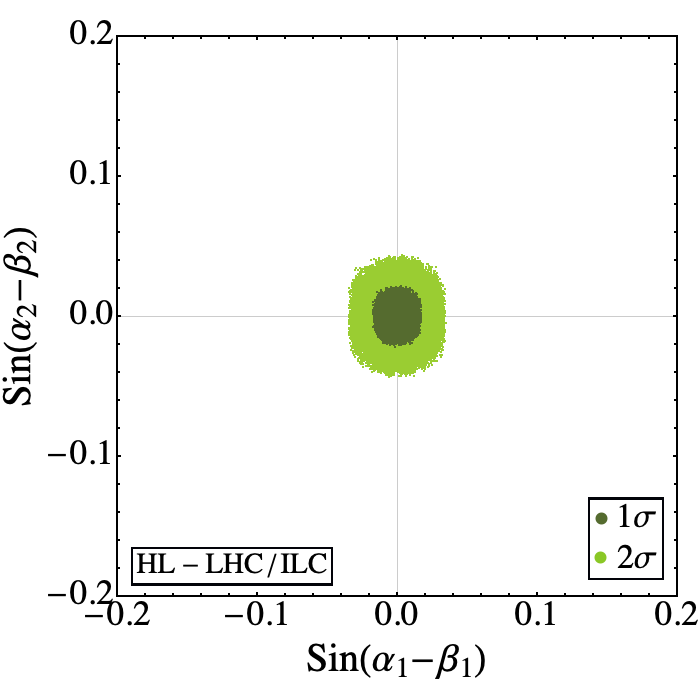}}
	\end{minipage}
	\caption{\em Allowed region in the $\sin(\alpha_1-\beta_1)$-$\sin(\alpha_2-\beta_2)$ plane from the current (left panel) \cite{ATLAS:2018doi} 
		and the future (right panel)~\cite{Fujii:2015jha} measurements of $\kappa_x$. 
		The darker and lighter shades represent $1\sigma$ and $2\sigma$ allowed
		regions respectively. While extracting bound using the projected accuracies at the HL-LHC and the ILC, the central
		values for all the $\kappa_x$ are assumed to be unity, {\it i.e.}, consistent with the SM.}
\label{f:allowedregion}
\end{figure} 

To provide a quantitative estimate of how close we need to be to the alignment limit, we perform a random scan
over the following parameters:
\begin{eqnarray}
\alpha_1\,, \alpha_2 \in \left[ - \frac{\pi}{2} , \frac{\pi}{2} \right] \,; \qquad \beta_1\,, \beta_2 \in \left[ 0 , \frac{\pi}{2} \right] \,.
\end{eqnarray}

The set of points that successfully negotiate the experimental constraints on $\kappa_x$ have been plotted in the $\sin(\alpha_1-\beta_1)$-$\sin(\alpha_2- \beta_2)$ plane as shown
in Fig.~\ref{f:allowedregion}. From Fig.~\ref{f:allowedregion}, it is evident that as the Higgs data converges towards the SM expectations with increasing accuracy, we are pushed closer to the alignment limit.

To illustrate further the usefulness of the alignment conditions
given in \Eqn{e:alfinal}, let us start by writing the scalar
potential for the $Z_3$-symmetric 3HDM\cite{Bento:2017eti}:
\begin{eqnarray}
V &=& m_{11}^2(\phi_1^\dagger \phi_1) +  m_{22}^2(\phi_2^\dagger 
\phi_2) +  m_{33}^2 (\phi_3^\dagger \phi_3)  + \lambda_1(\phi_1^\dagger 
\phi_1)^2 + \lambda_2(\phi_2^\dagger \phi_2)^2 + 
\lambda_3(\phi_3^\dagger \phi_3)^2  \nonumber \\ 
&& + \lambda_4 (\phi_1^\dagger \phi_1)( \phi_2^\dagger \phi_2) + \lambda_5 (\phi_1^\dagger \phi_1)( \phi_3^\dagger \phi_3) + 
 \lambda_6 (\phi_2^\dagger \phi_2)( \phi_3^\dagger \phi_3) \nonumber \\
&& + \lambda_7 (\phi_1^\dagger \phi_2)( \phi_2^\dagger \phi_1) +
\lambda_8 (\phi_1^\dagger \phi_3)( \phi_3^\dagger \phi_1)+ \lambda_9
(\phi_2^\dagger \phi_3)( \phi_3^\dagger \phi_2) \nonumber \\
&& + \left[\lambda_{10} (\phi_1^\dagger \phi_2)( \phi_1^\dagger \phi_3) 
+ \lambda_{11} (\phi_1^\dagger \phi_2)( \phi_3^\dagger \phi_2)+
\lambda_{12} (\phi_1^\dagger \phi_3)( \phi_2^\dagger \phi_3) + {\rm h.c.} \right] \,.
\label{e:potential}
\end{eqnarray}
We assume that all the parameters in the scalar potential are real.
Now let us ask how one can find a suitable set of values for the
potential parameters, which is compatible with a $125$~GeV Higgs
having SM-like properties. The usual procedure involves a random
scan over the parameter space and selecting those which satisfy
the condition of an SM-like Higgs within the experimental
uncertainties. Needless to say that such a brute force method
is quite inefficient. Therefore, any alternative approach that
can offer a more elegant strategy to recover an SM-like Higgs boson
from the 3HDM scalar potential will be beneficial for future
analyses of 3HDMs in view of the Higgs data.

To this end, we note that the scalar potential of \Eqn{e:potential}
contains fifteen parameters. Among them, the bilinear parameters
$m_{11}^2$, $m_{22}^2$ and $m_{33}^2$ can be traded for the three
VEVs, $v_1$, $v_2$ and $v_3$ or equivalently $v$, $\tan\beta_1$
and $\tan\beta_2$.The remaining twelve quartic couplings can be
exchanged for seven physical masses (three CP-even scalars,
two CP-odd scalars and two pairs of charged scalars) and five
mixing angles (three in the CP-even sector, one in the CP-odd
sector and one in the charged scalar sector). To demonstrate this
explicitly, let us examine the potential of \Eqn{e:potential}
in some more detail.

We start by using the minimization conditions to trade the
bilinear parameters in favor of the VEVs as follows:
\begin{subequations}
    \begin{eqnarray}
    m_{11}^2 &=& -\lambda_1 v_1^2 -\frac{1}{2}\left\{
    (\lambda_4+\lambda_7)v_2^2 +(\lambda_5+\lambda_8)v_3^2
    +2\lambda_{10}v_2 v_3\right\} -\frac{v_2v_3}{2v_1}
    \left(\lambda_{11}v_2 +\lambda_{12}v_3 \right) \,, \\
    m_{22}^2 &=& -\lambda_2 v_2^2 -\frac{1}{2}\left\{
    (\lambda_4+\lambda_7)v_1^2 +(\lambda_6+\lambda_9)v_3^2
    +2\lambda_{11}v_1 v_3\right\} -\frac{v_1v_3}{2v_2}
    \left(\lambda_{10}v_1 +\lambda_{12}v_3 \right) \,, \\
    m_{33}^2 &=& -\lambda_3 v_3^2 -\frac{1}{2}\left\{
    (\lambda_5+\lambda_8)v_1^2 +(\lambda_6+\lambda_9)v_2^2
    +2\lambda_{12}v_1 v_2\right\} -\frac{v_1v_2}{2v_3}
    \left(\lambda_{10}v_1 +\lambda_{11}v_2 \right) \,.
    \end{eqnarray}
    \label{e:bilinears}
\end{subequations}
Now let us investigate the mass matrices in different sectors.


\subsection{CP-odd scalar sector} 
The mass term for the  pseudoscalar sector can be extracted from the scalar potential as,
		\begin{eqnarray}\label{e:cpoddmassmat}
		V_{P}^{\rm mass} = \begin{pmatrix}
		z_1 & z_2 & z_3
		\end{pmatrix} \, \frac{{\cal M}_P^2}{2} \, \begin{pmatrix}
		z_1\\  z_2\\ z_3\\
		\end{pmatrix} \,,
		\end{eqnarray}
 where ${\cal M}_{P}^2$ is the $3\times3$ mass matrix which can be block diagonalized as follows:\footnote{Such a block diagonalization
     in the pseudoscalar and the charged scalar sector is a general
     property of CP conserving 3HDMs.}
%
	\begin{subequations}
		\begin{eqnarray}\label{e:mp2x2}
		({\cal B}_{P})^2 \equiv {\cal O}_{\beta} \cdot {\cal M}_{P}^2 \cdot {\cal O}_{\beta}^{T} &=& \begin{pmatrix}
		0 & 0 & 0 \\
		0 & {({\cal B}_P^2)}_{22}  & {({\cal B}_P^2)}_{23} \\
		0 & {({\cal B}_P^2)}_{23} & {({\cal B}_P^2)}_{33} \\
		\end{pmatrix} \,.
		\end{eqnarray}
The elements of ${{\cal B}_P^2}$ are given by,
		\begin{eqnarray}
		{({\cal B}_P^2)}_{22} &=& -\frac{v_3}{2v_1 v_2\left(v_1^2+v_2^2\right)}  \left[ \lambda_{10} v_1\left(v_1^2+2 v_2^2\right)^2+\lambda_{11} v_2\left(2 v_1^2+v_2^2\right)^2+\lambda_{12} v_3 \left(v_1^2-v_2^2\right)^2\right] \,,
		\\
		{({\cal B}_P^2)}_{23}  &=&\frac{v }{2 \left(v_1^2+v_2^2\right)} \left[ -\lambda_{10} v_1 \left(v_1^2 + 2 v_2^2\right)+ \lambda_{11}v_2 \left(2 v_1^2 +v_2^2\right)+ 2\lambda_{12} v_3 \left(v_1^2-v_2^2\right) \right] \,,\\ 
		{({\cal B}_P^2)}_{33} &=&  -\frac{v^2}{2 v_3\left(v_1^2+v_2^2\right)}  \left[\lambda_{10}v_1^2 v_2+ \lambda_{11} v_1 v_2^2+ 4\lambda_{12} v_1 v_2v_3\right] \,.
		\end{eqnarray}\label{e:BP2}
	\end{subequations}

The matrix ${{\cal B}_P^2}$ can be fully diagonalized using a orthogonal transformation as follows:
\begin{subequations}
	\begin{eqnarray}
	{\cal O}_{\gamma_1} \cdot ({\cal B}_{P})^2 \cdot {\cal O}_{\gamma_1}^T &=& \begin{pmatrix}
	0 & 0 & 0 \\
	0 & m_{A1}^2 & 0 \\
	0 & 0 & m_{A2}^2 \\
	\end{pmatrix}\,,\label{e:PSrot} 
	\end{eqnarray}
where,
	\begin{eqnarray}
{\cal O}_{\gamma_1} = 
\begin{pmatrix}
1 & 0 & 0 \\ 
0 & \cos\gamma_1 & -\sin\gamma_1 \\ 
0 & \sin\gamma_1 & \cos\gamma_1 \end{pmatrix}  \label{e:Ogamma1} \,.
\end{eqnarray}
\end{subequations}
This last step of diagonalization will entail the following relations
\begin{subequations}\label{e:mAtoBP2}
\begin{eqnarray}
m^2_{A1} \cos^2 \gamma_1 +  m^2_{A2} \sin^2 \gamma_1 &=& {({\cal B}_P^2)}_{22}   \,, \\
\cos \gamma_1 \sin\gamma_1 (m^2_{A2} - m^2_{A1})  &=&	{({\cal B}_P^2)}_{23} \,, \\
m^2_{A1} \sin^2 \gamma_1 +  m^2_{A2} \cos^2 \gamma_1 &=& {({\cal B}_P^2)}_{33}  \,.
\end{eqnarray}
\end{subequations}
Using Eq.~(\ref{e:BP2}), we can now invert Eq.~(\ref{e:mAtoBP2}) to solve for $\lambda_{10}\,, \lambda_{11}$ and $\lambda_{12}$ as follows:
\begin{subequations}\label{e:lam12}
	\begin{eqnarray}
	 \lambda_{10} &=&
	 \frac{2m_{A_1}^2}{9 v^2} \left[\frac{s_{2\gamma_1}}{c_{\beta_1} c_{\beta_2}}-\frac{2s_{\beta_1}  c^2_{\gamma_1}}{s_{\beta_2} c_{\beta_2}}+\frac{s_{3\beta_1} s_{\gamma_1} c_{\gamma_1}}{s_{\beta_1} c_{\beta_1}c_{\beta_2}}+\tan\beta_2 s^2_{\gamma_1} \left\{\frac{\tan\beta_1}{c_{\beta_1}}-2c_{\beta_1} \cot\beta_1\right\}\right]\nonumber \\
	 &&-\frac{m_{A_2}^2}{9 v^2} \left[(2 c_{2\beta_1}+3)\frac{s_{2\gamma_1}}{ c_{\beta_1} c_{\beta_2} }+4\frac{s_{\beta_1}  s^2_{\gamma_1}}{s_{\beta_2} c_{\beta_2}}-2 \tan\beta_2 c^2_{\gamma_1} \left\{\frac{\tan\beta_1}{c_{\beta_1}}-2 c_{\beta_1}\cot \beta_1\right\}\right]\,, \\
	\lambda_{11} &=& \frac{m_{A1}^2}{9 v^2} \left[ - \frac{4 c_{\beta_1} c_{\gamma_1}^2}{s_{\beta_2}c_{\beta_2}} + \frac{(-3 + 2 c_{2\beta_1})}{s_{\beta_1}c_{\beta_2}} s_{2\gamma_1} + 2 \left(\cot^4 \beta_1 + \cot^2 \beta_1-2\right)s_{\beta_1}s_{\gamma_1}^2 \tan \beta_1 \tan \beta_2 \right] \nonumber \\
	&&+\frac{m_{A2}^2}{9 v^2}\left[ - \frac{4 c_{\beta_1} s_{\gamma_1}^2}{s_{\beta_2}c_{\beta_2}} + \frac{(5 + \cot^2\beta_1)}{c_{\beta_2}} s_{2\gamma_1}s_{\beta_1} + 2 \left(\cot^4\beta_1 + \cot^2 \beta_1-2\right)s_{\beta_1}c_{\gamma_1}^2 \tan \beta_1 \tan \beta_2 \right]\,, \\
	\lambda_{12}  &=& \frac{m_{A1}^2}{36 v^2} \left[\frac{4 s_{2\beta_1}  c^2_{\gamma_1}}{s ^2_{\beta_2}}-\frac{4 c_{2\beta_1}  s_{2\gamma_1}}{s_{\beta_2}}+(c_{4\beta_1}-17) 
	\frac{s^2_{\gamma_1}}{s_{\beta_1} c_{\beta_1}}\right] \nonumber \\
	&&+\frac{m_{A2}^2}{36 v^2} \left[ \frac{4s_{2\beta_1} s^2_{\gamma_1}}{ s^2_{\beta_2}}+\frac{4c_{2\beta_1} s_{2\gamma_1}}{s_{\beta_2}} +(c_{4\beta_1}-17) \frac{ c^2_{\gamma_1}}{s_{\beta_1} c_{\beta_1}}\right] \,,
	\end{eqnarray}
\end{subequations}
where $s_x$ and $c_x$ stand for $\sin x$ and $\cos x$ respectively.


\subsection{Charged scalar sector} 
Similar to the pseudoscalar case, the $3\times3$ charged sector mass matrix ${\cal M}_{C}^2$ can also be block diagonalized as:
\begin{subequations}\label{e:BC2}
	\begin{eqnarray}
	({\cal B}_{C})^2 \equiv {\cal O}_{\beta} \cdot {\cal M}_{C}^2 \cdot {\cal O}_{\beta}^{T} &=& \begin{pmatrix}
	0 & 0 & 0 \\
	0 & {({\cal B}_C^2)}_{22}  & {({\cal B}_C^2)}_{23} \\
	0 & {({\cal B}_C^2)}_{23} & {({\cal B}_C^2)}_{33} \\
	\end{pmatrix} \,.
	\end{eqnarray}
	where,
	\begin{eqnarray}
	{({\cal B}_C^2)}_{22} &=& -\frac{1}{2(v_1^2+v_2^2)} \bigg[\lambda_{10}\frac{v_3}{v_2} \left(\left(v_1^2+v_2^2\right)^2+v_2^4\right) + \lambda_{11}\frac{v_3}{v_1} \left(\left(v_1^2+v_2^2\right)^2+v_1^4\right) + \lambda_{12}\frac{v_3^2}{v_1 v_2} \left(v_1^4+v_2^4\right)\nonumber \\
	&& + \lambda_7 \left(v_1^2+v_2^2\right)^2+ \lambda_8 v_2^2 v_3^2+ \lambda_9 v_1^2 v_3^2\bigg] \,, \\
	{({\cal B}_C^2)}_{23}  &=& \frac{v}{2(v_1^2+v_2^2)} \left[- v_1 v_2^2 \lambda_{10} + \lambda_{11} v_1^2 v_2 + \lambda_{12} v_3(v_1^2 - v_2^2)- \lambda_8 v_1 v_2 v_3 + \lambda_9 v_1 v_2 v_3 \right] \,, \\
	{({\cal B}_C^2)}_{33} &=& -\frac{v^2}{2(v_1^2+v_2^2)} \left[\frac{v_1^2 v_2}{v_3}\lambda_{10} + \lambda_{11} \frac{v_1 v_2^2}{v_3} + 2 v_1 v_2 \lambda_{12} + \lambda_8 v_1^2 + \lambda_9 v_2^2\right] \,.
	\end{eqnarray}
\end{subequations}
We completely diagonalize the charged scalar mass matrix as,
	\begin{subequations}\label{e:BCrot}
	\begin{eqnarray}
	{\cal O}_{\gamma_2} \cdot ({\cal B}_{C})^2 \cdot {\cal O}_{\gamma_2}^T = \left(\begin{array}{ccc}
	0 & 0 & 0 \\
	0 & m_{C1}^2 & 0 \\
	0 & 0 & m_{C2}^2 \\
	\end{array}\right)\,,
	\end{eqnarray}
	where,
	\begin{eqnarray}
{\cal O}_{\gamma_2} = 
\begin{pmatrix}
1 & 0 & 0 \\ 
0 & \cos\gamma_2 & -\sin\gamma_2 \\ 
0 & \sin\gamma_2 & \cos\gamma_2 \end{pmatrix}  \label{e:Ogamma2} \,.
\end{eqnarray}
\end{subequations}

Thus, we will have the following relations:
\begin{subequations}\label{e:mctoapbpcp}
\begin{eqnarray}
m^2_{C1} \cos^2 \gamma_2 +  m^2_{C2} \sin^2 \gamma_2 &=& {({\cal B}_C^2)}_{22}\,, \\
\cos \gamma_2 \sin\gamma_2 (m^2_{C2} - m^2_{C1}) &=&  {({\cal B}_C^2)}_{23}\,, \\
m^2_{C1} \sin^2 \gamma_2 +  m^2_{C2} \cos^2 \gamma_2 &=&  {({\cal B}_C^2)}_{33}\,.
\end{eqnarray}
\end{subequations}
These equations in conjunction with Eq.~(\ref{e:BC2}) will enable us to solve for $\lambda_{7},\lambda_8$, and $\lambda_9$ as given below:
\begin{subequations} \label{e:lam789}
	\begin{eqnarray}
 \lambda_7  &=& 
 \frac{\left(m_{C1}^2 - m_{C2}^2\right)}{2 v^2} \left[(-3 + c_{2\beta_2})\frac{c_{2\gamma_2}}{c_{\beta_2}^2} + \frac{4\tan \beta_2}{\tan 2\beta_1} \frac{ s_{2\gamma_2}}{c_{\beta_2}}\right] -
  \frac{\left(m_{C1}^2 + m_{C2}^2\right)}{v^2}
  -\lambda_{10}\frac{\tan\beta_2}{ s_{\beta_1} } -\lambda_{11} \frac{\tan\beta_2}{c_{\beta_1}} \,, \\
\lambda_8  &=&\frac{m_{C1}^2}{v^2}\left(-2s^2_{\gamma_2}+\tan\beta_1 \frac{s_{2\gamma_2} }{s_{\beta_2} }\right)
-\frac{m_{C2}^2}{v^2}\left(2c^2_{\gamma_2}+\tan\beta_1  \frac{ s_{2\gamma_2}}{s_{\beta_2}}\right) 
- \lambda_{10} s_{\beta_1} \cot\beta_2 -\lambda_{12} \tan\beta_1\,, \\
\lambda_9  &=& -\frac{m_{C1}^2}{v^2}\left(2s^2_{\gamma_2}+\cot\beta_1 \frac{s_{2\gamma_2} }{ s_{\beta_2}}\right)
+\frac{m_{C2}^2}{v^2}\left(-2c^2_{\gamma_2}+\cot\beta_1  \frac{s_{2\gamma_2} }{s_{\beta_2}}\right) 
-\lambda_{11} c_{\beta_1} \cot\beta_2 -\lambda_{12} \cot\beta_1\,.
	\end{eqnarray}
\end{subequations}
where, the other three couplings $(\lambda_{10},\lambda_{11 }~\& ~\lambda_{12})$ can be replaced using Eq.~(\ref{e:lam12}). 


\subsection{CP-even scalar sector}
The mass terms in the neutral scalar sector can be extracted from the potential as,
	\begin{subequations} 
	\begin{eqnarray}\label{e:neutralmassmat}
	V_{S}^{\rm mass} =\begin{pmatrix}
	h_1 & h_2 & h_3\\
	\end{pmatrix} \frac{{\cal M}_S^2}{2} \begin{pmatrix}
	h_1\\  h_2\\ h_3\\
	\end{pmatrix} \,,
	\end{eqnarray}
where, ${\cal M}_S^2$ is the $3\times3$ symmetric
mass matrix whose elements are given by,
	\begin{eqnarray}
	{({\cal M}_S^2)}_{11} &=& 2 v_1^2 \lambda_1 - \frac{v_2 v_3 \left(v_2 \lambda_{11} + v_3 \lambda_{12}\right)}{2 v_1}\,, \\
	{({\cal M}_S^2)}_{12} &=&  v_1\left(v_2 (\lambda_{7}+ \lambda_4)+ v_3 \lambda_{10}\right) + \frac{v_3}{2} \left(2 v_2 \lambda_{11} + v_3 \lambda_{12}\right) \,, \\
	{({\cal M}_S^2)}_{13} &=&  v_1\left(v_3 (\lambda_8+ \lambda_5)+ v_2 \lambda_{10}\right)+ \frac{v_2}{2} \left( v_2 \lambda_{11} +2 v_3 \lambda_{12}\right) \,, \\
	{({\cal M}_S^2)}_{22} &=& 2 v_2^2 \lambda_2 - \frac{v_1 v_3 \left(v_1 \lambda_{10} + v_3 \lambda_{12}\right)}{2 v_2} \,, \\ 
	{({\cal M}_S^2)}_{23} &=& v_3\left( v_2 (\lambda_{6}+ \lambda_9)+ v_1 \lambda_{12}\right) + \frac{v_1}{2} \left(2 v_2 \lambda_{11} + v_1 \lambda_{10}\right)  \,, \\
	{({\cal M}_S^2)}_{33} &=& 2 v_3^2 \lambda_3 - \frac{v_1 v_2 \left(v_1 \lambda_{10} + v_2 \lambda_{11}\right)}{2 v_3} \,.
	\end{eqnarray}\label{e:mselement}
\end{subequations}
 This mass matrix should be diagonalized via the following orthogonal transformation
 \begin{eqnarray}\label{e:msdiag}
{\cal O}_\alpha \cdot {\cal M}_{S}^2 \cdot {\cal O}_\alpha^T &\equiv& \begin{pmatrix}
 m_h^2 & 0 & 0 \\
 0& m_{H1}^2 & 0 \\
 0 & 0 & m_{H2}^2 \\
 \end{pmatrix}  \,,
\end{eqnarray}
where, ${\cal O}_\alpha$ has already been defined in Eq.~(\ref{e:Oa}).
Inverting the above Eq.~(\ref{e:msdiag}), we get,
\begin{eqnarray}\label{e:MS}
  {\cal M}_{S}^2  &\equiv& {\cal O}_\alpha^T  \cdot \begin{pmatrix}
m_h^2 & 0 & 0 \\
0& m_{H1}^2 & 0 \\
0 & 0 & m_{H2}^2 \\
\end{pmatrix} \cdot {\cal O}_\alpha\,,
\end{eqnarray}
which enables us to solve for the remaining six lambdas
as follows:
\begin{subequations}
	\label{e:lam1to6}
	\begin{eqnarray}
	\lambda_1  &=& \frac{m_h^2}{2 v^2} \frac{c^2_{\alpha_1} c^2_{\alpha_2} }{c^2_{\beta_1} c^2_{\beta_2}}
	+ \frac{m_{H_1}^2}{2 v^2 c^2_{\beta_1} c^2_{\beta_2} }  \left( c_{\alpha_1} s_{\alpha_2}  s_{\alpha_3} + s_{\alpha_1} c_{\alpha_3}  \right)^2 
	+ \frac{m_{H_2}^2}{2 v^2 c^2_{\beta_1} c^2_{\beta_2}} \left(c_{\alpha_1}   s_{\alpha_2}  c_{\alpha_3} - s_{\alpha_1} s_{\alpha_3}  \right)^2 \nonumber \\
	&&+\frac{\tan\beta_1  \tan \beta_2}{4c^2_{\beta_1}} \left( \lambda_{11} s_{\beta_1} + \lambda_{12} \tan\beta_2  \right) \,, \\
	\lambda_2  &=& \frac{m_h^2}{2 v^2}\frac{s^2_{\alpha_1} c^2_{\alpha_2}  } {s^2_{\beta_1}c^2_{\beta_2}  }
	+\frac{m_{H_1}^2}{2 v^2 s^2_{\beta_1} c^2_{\beta_2}} \left(c_{\alpha_1} c_{\alpha_3} - s_{\alpha_1}s_{\alpha_2} s_{\alpha_3} \right)^2
	+\frac{m_{H_2}^2}{2 v^2 s^2_{\beta_1} c^2_{\beta_2}} \left(c_{\alpha_1} s_{\alpha_3} + s_{\alpha_1}s_{\alpha_2} c_{\alpha_3} \right)^2\nonumber \\
	&&+\frac{ \tan\beta_2}{4s^2_{\beta_1}\tan \beta_1 } \left( \lambda_{10} c_{\beta_1}+ \lambda_{12} \tan\beta_2\right) \,, \\
	\lambda_3  &=& \frac{m_h^2}{2v^2} \frac{s^2_{\alpha_2}}{s^2_{\beta_2}}
	+\frac{m_{H_1}^2 c^2_{\alpha_2} s^2_{\alpha_3}}{2v^2 s^2_{\beta_2}}  
	+\frac{m_{H_2}^2 c^2_{\alpha_2} c^2_{\alpha_3}}{2v^2 s^2_{\beta_2}} 
	+\frac{s_{2\beta_1}  }{8\tan ^3\beta_2}  \left( \lambda_{10}c_{\beta_1} + \lambda_{11}s_{\beta_1}  \right)
	 \,,\\
	\lambda_4  &=& 
	\frac{1}{4v^2 s_{2\beta_1} c^2_{\beta_2}}\left[\left(m_{H_1}^2-m_{H_2}^2\right) \left\{(-3 + c_{2\alpha_2})s_{2\alpha_1}c_{2\alpha_3} - 4 c_{2\alpha_1} s_{\alpha_2} s_{2\alpha_3}\right\}
	-2\left(m_{H_1}^2+m_{H_2}^2\right)s_{2\alpha_1} c^2_{\alpha_2}  \right] \nonumber \\
	&&+\frac{m_h^2}{v^2}\frac{  s_{2\alpha_1}c^2_{\alpha_2}}{s_{2\beta_1} c^2_{\beta_2}} 
	- \frac{\tan\beta_2}{ s_{2\beta_1}}\left(2\lambda_{10}c_{\beta_1} + 2\lambda_{11}s_{\beta_1} + \lambda_{12} \tan\beta_2 \right)-\lambda_{7}   \,, \\
	\lambda_5 &=& \frac{m_h^2}{ v^2 } \frac{c_{\alpha_1} s_{2\alpha_2}}{c_{\beta_1} s_{2\beta_2}}
	-\frac{m_{H_1}^2}{v^2c_{\beta_1} s_{2\beta_2}} \left(c_{\alpha_1} s_{2\alpha_2} s^2_{\alpha_3}  + s_{\alpha_1}  c_{\alpha_2}s_{2\alpha_3} \right)  
    +\frac{m_{H_2}^2}{v^2c_{\beta_1} s_{2\beta_2}} \left( s_{\alpha_1} c_{\alpha_2}  s_{2\alpha_3}-c_{\alpha_1} s_{2\alpha_2} c^2_{\alpha_3}\right)  \nonumber \\
	&&- \frac{s_{\beta_1}}{2\tan\beta_2} \left( 2\lambda_{10} + \lambda_{11}\tan\beta_1\right) -\lambda_{12} \tan\beta_1- \lambda_8 \,, \\
	\lambda_6  &=&  \frac{ m_h^2 }{ v^2}\frac{s_{\alpha_1}s_{2\alpha_2} } {s_{\beta_1} s_{2\beta_2}} 
	+ \frac{m_{H_1}^2 }{v^2}\frac{c_{\alpha_2}}{s_{\beta_1} s_{2\beta_2}} \left(-2s_{\alpha_1} s_{\alpha_2} s^2_{\alpha_3}  + c_{\alpha_1} s_{2\alpha_3} \right)  
	- \frac{m_{H_2}^2}{v^2} \frac{c_{\alpha_2}} {s_{\beta_1}s_{2\beta_2} } \left(2s_{\alpha_1} s_{\alpha_2} c^2_{\alpha_3}  + c_{\alpha_1} s_{2\alpha_3} \right)\nonumber \\
	&& - \frac{c_{\beta_1}}{2\tan\beta_2} \left(\lambda_{10} \cot\beta_1  +2 \lambda_{11}\right) -\lambda_{12} \cot\beta_1-\lambda_{9}\,.
	\end{eqnarray} 
\end{subequations}

\subsection{Implementing the alignment limit}
With Eqs.~(\ref{e:lam12}), (\ref{e:lam789}) and (\ref{e:lam1to6}) in hand, we can now go back to the problem of finding a set of lambdas consistent with a 125 GeV SM-like Higgs boson.
This can now be achieved quite simply by putting $m_h = 125~\rm GeV$, $\alpha_1=\beta_1$ and $\alpha_2=\beta_2$ in  Eqs.~(\ref{e:lam12}), (\ref{e:lam789}) and (\ref{e:lam1to6}). Moreover,
deviations from the exact alignment limit can also be parametrized rather conveniently. Defining $\sin (\alpha_1-\beta_1) = \delta_1$ and $\sin (\alpha_2-\beta_2) = \delta_2$, one can use,
\begin{eqnarray}
\alpha_1 = \sin^{-1} (\delta_1) + \beta_1 \,; \qquad \alpha_2 = \sin^{-1} (\delta_2) + \beta_2 \,,
\end{eqnarray}
 to extract $\alpha_1$ and $\alpha_2$ and then putting them back in Eqs.~(\ref{e:lam12}), (\ref{e:lam789}) and (\ref{e:lam1to6}) to compute the lambdas. Thus, the final result can be obtained in terms of the deviations, $\delta_1$ and $\delta_2$ with $\delta_1=\delta_2=0$ characterizing the exact alignment limit.

Before we conclude, it should be noted that Eqs.~(\ref{e:lam12}), (\ref{e:lam789}) and (\ref{e:lam1to6}) allow us to express the
scalar self couplings in terms of the physical parameters. To
illustrate, one can write the charged Higgs trilinear couplings
with the SM-like Higgs scalar as follows:
\begin{eqnarray}
    {\mathscr L}_{H_i^+H_i^-h} = g_{H_i^+H_i^-h} \, H_i^+H_i^-h \,,
    \qquad (i=1,2) \,.
\end{eqnarray}
Using Eqs.~(\ref{e:lam12}), (\ref{e:lam789}) and (\ref{e:lam1to6})
one can then calculate
\begin{eqnarray}
g_{H_i^+H_i^-h} = -\frac{1}{v}\left(m_h^2 + 2m_{Ci}^2\right) = -\frac{g m_{Ci}^2}{M_W}\left(1+ \frac{m_h^2}{2m_{Ci}^2}\right) \,,
\qquad (i=1,2) \,,
\end{eqnarray}
in the alignment limit, $M_W$ being the mass of the W-boson. Thus, non negligible contributions to
decay processes like $h\to \gamma\gamma$ can arise even from
super heavy charged scalars, which will strongly constrain
the $Z_3$-symmetric 3HDM\cite{Bhattacharyya:2014oka}. Terms that
break the $Z_3$ symmetry softly should be included in the scalar
potential to avoid such strong constraints.
 

\section{Summary}
\label{s:summary}
To summarize, we have presented a recipe for recovering a SM-like Higgs boson with a mass 125 GeV from the 3HDM scalar spectrum.
We have advocated a suitable parametrization in which such an alignment limit looks very similar to the corresponding limit in 2HDM case.
Using a $Z_3$ symmetric 3HDM as an example, we have demonstrated that our alignment conditions are simple enough to be easily implemented 
in a practical scenario which is a clear  upshot of our analysis. Although the topic of 3HDMs is well-trodden in the literature, the existence of an
alignment limit described by such simple analytic conditions does not appear to be a widespread knowledge. Given the growing interest of the community in the
topic of multi Higgs-doublet models, number of studies on the constraints faced by such models from the Higgs data is expected to rise in the coming years. Thus,
the fact that our analysis provides a way to efficiently implement the alignment limit in case of a CP-conserving 3HDM, makes our results quite timely and relevant.

\section*{Acknowledgements}
The work of DD has been supported by the Swedish Research Council, contract number 2016-05996. DD gratefully acknowledges the
warm hospitality of
Kavli IPMU where part of this work has been completed. The work of IS was supported by World Premier International Research Center Initiative (WPI), MEXT, Japan.
%


\bibliographystyle{JHEP}
\bibliography{ref.bib}
\end{document}